# The Dual-Loop Model of Psychophysiological Regulation: A Framework for Psychological Breakthrough and State Transition

Xueqng Deng

Universit Antwerpen

Faculty of Sciences - Department of Biology

Campus Drie Eiken - C1.31

Universiteistplein - 2610 Antwerpen - België

xueqing.deng@student.uantwerpen.be

xueqing.deng2023@outlook.com

**Abstract**

In real life, psychological and physiological states rarely change along a single dimension. Through self-tracking and discussions with clinicians, I have come to recognise with increasing clarity that sleep patterns, autonomic arousal, bodily sensations, and cognitive load are in constant interaction. Existing models often fail to capture this complexity. Many theoretical frameworks continue to analyse these elements in isolation, making it difficult to explain sudden changes reported by individuals—such as abrupt spikes in anxiety, sudden drops in dissociation, or even moments of heightened alertness.

To bridge this gap, this study proposes the Dual-Loop Sleep-Cognitive Regulation Framework (DLSSC-F). This model integrates four dimensions—sleep, autonomic nervous system, somatic perception, and cognitive load—into a standardised shared system using Z-scores. Analysing these interactions reveals a key concept: the 'psychological tipping point'. Identifiable and measurable state transitions occur when two regulatory loops (Loop 1: somatic load; Circuit 2: Cognitive interpretative load) begin to influence each other non-linearly, an identifiable and measurable state transition occurs.

The mathematical modelling employed herein does not replace clinical or subjective narratives, but rather provides a structural framework for these rapid transitions and elucidates why bodily-driven and cognitively-driven changes manifest differently. The objective is to build a conceptual bridge between physiological signals and lived experience, laying the groundwork for dynamic modelling and future case analyses.

**Keyword：dual-loop regulation; psychophysiology; psychological breakthrough; state transition; sleep debt; autonomic arousal; somatic perception; cognitive load**

## 1. Introduction

Subjectively, people tend to believe that emotional changes synchronise with physical reactions, yet the reality is quite the opposite. More often than not, individuals only become aware of their nervousness when their heart rate accelerates – meaning the bodily response precedes the psychological emotion. Conversely, when psychological equilibrium is restored, certain physical reactions may persist without stabilising. Stress appears manageable until it suddenly spirals out of control; an individual may maintain composure for hours only to experience a panic attack without warning; sustained effort may not culminate in a gradual decline but instead result in abrupt collapse. Such patterns are commonplace in both clinical settings and everyday accounts.

Classic affect theories (e.g., Barrett, 2017; Damasio, 1994) and emerging interoceptive theories (Critchley & Garfinkel, 2017) both emphasise the close connection between bodily signals and the meanings they construct. Yet these perspectives struggle to explain: why do emotional shifts sometimes unfold continuously, while at other times reconstruct themselves abruptly or discontinuousl.

Traceable to James's (1884) research and expanded upon by contemporary computational theories (Seth & Friston, 2016), it is shown that bodily fluctuations frequently precede conscious emotional interpretation. This sheds light on the possibility that distinct components of emotional regulation may operate at differing speeds. Building upon this perspective, I propose that psychological regulation can be effectively conceptualised as the interaction between two circuits: a relatively rapid somatic loop (SL) capable of swiftly responding to disturbances in physiological states, and a slower cognitive loop (CL) that integrates these signals into narratives, evaluations, and meanings.

Despite significant advances in understanding mind-body coupling, two major challenges persist. Firstly, existing frameworks largely lack formal mathematical descriptions, failing to explicitly capture non-linear or discontinuous emotional shifts. Secondly, the abrupt transitions frequently observed by clinicians and individuals during stress responses, recovery processes, or therapeutic work remain without established terminology. To address these gaps, I propose the concept of "psychological transitions", specifically defined as:

the rapid reconfiguration of systemic states within the double-loop architecture when functional load and somatic-cognitive coupling jointly drive the system beyond its critical threshold.

This idea helps explain why dramatic emotional changes can occur even when the underlying pressures accumulate gradually. To make this construct precise, the present work introduces the DLSSC-F framework, which represents SL and CL using a pair of coupled differential equations. Functional Load (F)—capturing elements of task demand, responsibility, and cognitive–somatic effort—acts as a nonlinear integrator that links the two loops. This structure produces the kinds of features typical of fast–slow dynamical systems (Strogatz, 2015), including nullclines, critical points, and state transitions. Notably, these mathematical properties correspond closely with real-world phenomena such as panic onset, abrupt cognitive reframing, emotional breakthroughs, and rapid recovery after high stress (Kuppens et al., 2010; Hayes et al., 2020).

The aims of this paper are threefold: (1) to outline DLSSC-F as a formal theoretical model of somatic–cognitive regulation; (2) to examine its dynamic properties using a mathematical approach; and (3) to clarify why the somatic and cognitive cycles behave differently in the presence of continuous load. My broader goal is to offer a framework that can account for both gradual emotional drift and the sudden shifts that emerge in many psychological experiences.

## 2. Theoretical Background

Understanding human self-regulation requires an integrative account of how sleep, autonomic activity, somatic tension, cognitive load, and functional burden interact across multiple timescales. Historically, research has treated bodily processes and cognitive processes as independent systems, despite strong evidence that they form a tightly coupled regulatory structure[15]–[18].

## 2.1 Fragmentation in Existing Somatic and Cognitive Research

Somatic research has traditionally concentrated on domains such as sleep pressure and sleep debt, autonomic activation, heart-rate dynamics, and bodily discomfort including muscle tension. In contrast, cognitive research has emphasized rumination, attentional fragmentation, working-memory load, and meaning reconstruction. These two research traditions evolved largely independently due to methodological differences, measurement conventions, and disciplinary boundaries[19–21]. Consequently, although real humans function as integrated psychophysiological systems, the scientific literature historically treats somatic and cognitive processes as separable subsystems. This mismatch becomes especially apparent during rapid psychological transitions—such as sudden emotional breakthroughs, exhaustion crashes, or acute anxiety collapses—which clearly involve simultaneous shifts in both somatic states and cognitive dynamics[22–24]. These patterns highlight the need for a unified mathematical framework capable of capturing their coupled behavior.

## 2.2 Why Multi-Axis Integration Is Necessary

The four DLSSC-F axes—Sleep, Autonomic, Somatic, and Functional Load—must be integrated because they exhibit inherent nonlinear interdependence. Sleep loss amplifies somatic sensitivity, including muscle tension, pain, and fatigue[25–27], while increased somatic tension elevates cognitive load through intensified intrusions and rumination[28,29]. Rising cognitive load, in turn, heightens functional burden and subjective effort[30,31]. This increased functional burden then worsens sleep pressure, completing the fast–slow feedback loop[32]. Because each axis dynamically influences the others, they cannot be modeled as independent components.

## 2.3 The Somatic Loop (Fast System)

The Somatic Loop (SL) integrates sleep pressure, somatic tension, autonomic activation (particularly in robust HR metrics), and functional burden insofar as it influences physical effort. It represents the fast-reacting subsystem that shifts over hours to days. This loop accounts for rapid psychophysiological phenomena such as sudden fatigue, acute physical anxiety, and tension-induced emotional reactivity. These processes unfold quickly and generate cascading effects on cognition[33,34].

## 2.4 The Cognitive Loop (Slow System)

The Cognitive Loop (CL) integrates rumination, cognitive load, attentional fragmentation, meaning-making, and functional load as a form of appraisal. It captures slow, integrative psychological dynamics that evolve over days or weeks[35]–[37]. This loop governs conceptual reframing, narrative reorganization, long-term psychological transitions, and the reduction of rumination following breakthroughs. Its slower timescale reflects the accumulation and processing of information rather than immediate bodily states.

**2.5 Functional Load as the Cross-Loop Connector**

Functional Load (F) simultaneously belongs to both loops: in the Somatic Loop, it represents external task demand, while in the Cognitive Loop, it represents internal role evaluation. Thus, F(t) acts as the mathematical and conceptual bridge between body and mind. Removing F(t) would decouple the Somatic and Cognitive Loops, eliminate cross-loop feedback, destroy nonlinear dynamic behavior, and prevent the model from capturing real psychological transitions. Functional Load is therefore the keystone variable of the DLSSC-F framework.

**2.6 Why a Two-Loop System (Not One, Not Many)?**

DLSSC-F adopts exactly two loops because bodily processes form a natural cluster and cognitive processes form a natural cluster, with functional load providing the minimal necessary coupling. This structure ensures mathematical tractability while avoiding unnecessary model inflation; adding additional loops does not meaningfully increase explanatory power. The dual-loop design is the simplest system capable of reproducing fast–slow regulation, cross-loop feedback, bifurcations, and sudden reorganizational transitions[38]–[40].

### 3. Variable Space and Domain Definition

The DLSSC-F model relies on four observable daily variables recorded across a period of TTT days. For mathematical rigor, each variable must have a clearly defined domain, a truncation procedure for extreme values, and a baseline distribution for Z-score standardization. These definitions ensure that the four-axis model produces stable, interpretable metrics even in real-world sampling conditions[41]–[44].

Let:

$$r = 1, 2, \ldots, T$$

represent discrete days.

Let:

$$X_r \in \{S_r, R_r, SomaticAvg_r, FuncLoad_r\}$$

represent any raw variable before preprocessing.

### 3.1 Raw Variable Definitions

#### 3.1.1 Sleep Gap Sr

$$S_r \in \mathbb{R}$$

Defined as:

$$S_r = \text{Actual Sleep Duration (h)} - \text{Ideal Sleep Duration (h)}$$

Domain justification: sleep surplus and deficit strongly modulate autonomic and somatic systems[45–47].

#### 3.1.2 Autonomic Index Rr

$$R_r \in \mathbb{R}$$

This paper uses robust heart-rate–based metrics rather than HRV or EDA due to noise sensitivity[12,48–50].

Common components:

- mean HR

- HR day–night difference
- smoothed HR variability (non-spectral)

Autonomic activation plays a central role in stress, arousal, and somatic–cognitive coupling[51,52].

### 3.1.3 Somatic Average

$$SomaticAvg_r$$

$$SomaticAvg_r \in [0, 10]$$

Represents daily bodily tension, discomfort, and physical sensitivity, consistent with somatic symptom theory[53,54].

### 3.1.4 Functional Load FuncLoadr

$$FuncLoad_r \in [0, 10]$$

Captures task burden, role pressure, executive energy and is the cross-loop connector between bodily and cognitive systems[55]–[57].

## 3.2 Winsorization: Extreme-Value Truncation

DLSSC-F requires stable baselines. To prevent contamination by outliers, all variables undergo bounded winsorization[41]:

$$\tilde{X}_r = \begin{cases} L_X, & X_r < L_X \\ X_r, & L_X \leq X_r \leq U_X \\ U_X, & X_r > U_X \end{cases}$$

Thresholds:

- Sleep Gap: $L_S = -6\text{h}, U_S = +4\text{h}$
- SomaticAvg: naturally [0,10]
- Functional Load: naturally [0,10]
- Autonomic Index: use percentiles

$$L_R = P_{1\%}, \quad U_R = P_{99\%}$$

This preserves the distribution while preventing extreme artifacts from destabilizing baseline estimation.

### 3.3 Baseline Window B

$$\mathcal{B} = \{r_1, \ldots, r_n\} \subseteq \{1, \ldots, T\}$$

Represents a period of relative psychological–somatic stability, consistent with allostatic baseline frameworks[3],[45].

Baseline mean:

$$\mu_X = \frac{1}{n} \sum_{r \in \mathcal{B}} \tilde{X}_r$$

Baseline standard deviation:

$$\sigma_X = \sqrt{\frac{1}{n-1} \sum_{r \in \mathcal{B}} (\tilde{X}_r - \mu_X)^2}$$

Minimum variance safeguard[42]:

$$\sigma_X^* = \max(\sigma_X, \varepsilon)$$

### 3.4 Z-score Standardization

For any variable Xr

$$Z_X(r) = \frac{\tilde{X}_r - \mu_X}{\sigma_X^*}$$

Z-normalization enables cross-domain comparability of dimensions with different units (hours, HR metrics, 0–10 scales) and is consistent with psychophysiological modeling practice[42,43].

### 3.5 Summary of Domains

| Variable | Domain | Unit | Interpretation |
|---|---|---|---|
| Sr | R | hours | sleep deficit/surplus |
| Rr | R | HR-based index | autonomic activation |
| SomaticAvgr | [0,10] | score | somatic tension |
| FuncLoadr | [0,10] | score | functional demand |

These definitions provide the formal basis for the model's mathematical development.

## 4. Mathematical Formulation of the DLSSC-F Framework

The DLSSC-F model formalizes four psychophysiological axes—Sleep, Autonomic, Somatic, and Functional Load—into a standardized set of Z-scores and integrates them into a dynamical dual-loop system. This section presents the complete derivation of these components, including truncation, baseline estimation, Z-normalization, composite scoring, and PDE coupling[41-44].

| Category | Variable | Symbol | Definition | Unit | Valid Range |
|---|---|---|---|---|---|
| Sleep Axis | Sleep duration | $H_r$ | Total hours slept | hours | 0–24 |
| Sleep Axis | Sleep Gap | $S_r$ | Sleep deficit relative to 8 h | hours | −8 to +8 |
| Autonomic Axis | Morning RHR | $HR^{(m)}_r$ | Resting heart rate (morning) | bpm | 30–150 |
| Autonomic Axis | Noon RHR | $HR^{(n)}_r$ | Resting heart rate (noon) | bpm | 30–150 |
| Autonomic Axis | Evening RHR | $HR^{(e)}_r$ | Resting heart rate (evening) | bpm | 30–150 |
| Autonomic Axis | Daily mean RHR | $R_r$ | Mean of three RHR readings | bpm | 30–150 |
| Somatic Axis | Stomach discomfort | $D_r$ | GI discomfort rating | rating | 0–3 |

| Axis | Indicator | Symbol | Description | Unit | Range |
|---|---|---|---|---|---|
| Somatic Axis | Tension | $E_r$ | Muscular tension | rating | 0–3 |
| Somatic Axis | Heart discomfort | $F_r$ | Chest discomfort | rating | 0–3 |
| Somatic Axis | Cold/Numbness | $G_r$ | Coldness or numbness | rating | 0–3 |
| Somatic Axis | Subjective fatigue | $F^{fatigue}_r$ | Perceived fatigue | rating | 0–3 |
| Somatic Axis | Breathlessness | $I_r$ | Shortness of breath | rating | 0–3 |
| Somatic Axis | Fatigue interference (body) | $FI_r$ | Fatigue impact on body | rating | 0–3 |
| Somatic Axis | Somatic average load | $SomaticAvg_r$ | Mean of 8 somatic indicators | composite | 0–3 |
| Functional Axis | Focus minutes | $F^{minutes}_r$ | Minutes of sustained focus | minutes | 0–720 |
| Functional Axis | Task completion | $C^{task}_r$ | Daily task completion percentage | % | 0–100 |
| Functional Axis | Efficiency self-rating | $E^{eff}_r$ | Self-rated task efficiency | rating | 0–3 |
| Functional Axis | Fatigue interference (cognitive) | $FI^{cog}_r$ | Fatigue impact on cognition | rating | 0–3 |
| Functional Axis | Focus Score | $FS_r$ | Scaled focus (0–100) | % | 0–100 |
| Functional Axis | Task Score | $TS_r$ | Task performance | % | 0–100 |
| | | | (0–100) | | |
| Functional Axis | Efficiency Score | ES_r | Scaled efficiency (0–100) | % | 0–100 |
| Functional Axis | Fatigue Interference Score | FFS_r | Reverse-scored fatigue interference | % | 0–100 |
| Functional Axis | Functional Score | FuncScore_r | Mean of FS, TS, ES, FFS | % | 0–100 |
| Functional Axis | Functional Load | FuncLoad_r | Reverse-scaled functional score | % | 0–100 |

## 4.1. Raw Daily Variables

For each day rrr, the following primary observations are recorded:

### 4.1.1 Sleep

- Actual sleep duration (hours):

$$H_r \in \mathbb{R}_{\geq 0}$$

### 4.1.2 Somatic Symptoms (0–3 ratings)

$$D_r, E_r, F_r, G_r, F_r^{fatigue}, I_r \in \{0, 1, 2, 3\}$$

represent stomach discomfort, muscular tension, chest discomfort, cold/numbness, subjective fatigue, and shortness of breath.

### 4.1.3 Autonomic Activity

Daily resting heart rate:

$$HR_r^{(m)}, \ HR_r^{(n)}, \ HR_r^{(e)}$$

measured morning, noon, and evening.

### 4.1.4 Functional Performance

$$F_r^{minutes}, \ C_r^{task}, \ E_r^{eff}, \ FI_r$$

represent focus minutes, task completion (%), self-rated efficiency (0–3), and fatigue-interference (0–3).

### 4.2. Derived Raw Axis Variables

### 4.2.1 Sleep Gap

Deviation from an 8-hour ideal reference:

$$S_r := 8 - H_r$$

### 4.2.2 Mean Resting Heart Rate

$$R_r := \frac{HR_r^{(m)} + HR_r^{(n)} + HR_r^{(e)}}{3}$$

### 4.2.3 Somatic Average Load

Integrating sleep gap, six body sensations, and fatigue interference:

$$SomaticAvg_r := \frac{S_r + D_r + E_r + F_r + G_r + F_r^{fatigue} + I_r + FI_r}{8}$$

### 4.2.4 Functional Load

(a) Functional Sub-Scores mapped to 0–100 scale

Focus:

$$FS_r = 100 \cdot \min\left(1, \frac{F_r^{minutes}}{240}\right)$$

Task completion:

$$TS_r = C_r^{task}$$

Efficiency:

$$ES_r = 100 \cdot \frac{E_r^{eff}}{3}$$

Fatigue-interference (reverse-scored):

$$FFS_r = 100 \cdot \left(1 - \frac{FI_r}{3}\right)$$

(b) Functional Score (higher = better function)

$$FuncScore_r = \frac{FS_r + TS_r + ES_r + FFS_r}{4}$$

(c) Functional Load (higher = worse function)

$$FuncLoad_r := 100 - FuncScore_r$$

### 4.3. Baseline Normalization Parameters

A baseline period of 14 days (index set $\mathcal{A}$, size $N_A$) is used to compute mean and standard deviation for each axis:

$$\mu_X = \frac{1}{N_A} \sum_{r \in \mathcal{A}} X_r$$

where

$$\sigma_X = \sqrt{\frac{1}{N_A - 1} \sum_{r \in \mathcal{A}} (X_r - \mu_X)^2}$$

### 4.4. Standardized Axis Scores (z-scores)

For each day r:

#### 4.4.1 Sleep Axis

$$Z_{sleep}(r) = \frac{S_r - \mu_{sleep}}{\sigma_{sleep}}$$

#### 4.4.2 Autonomic Axis

$$Z_{auto}(r) = \frac{R_r - \mu_{auto}}{\sigma_{auto}}$$

### 4.4.3 Somatic Axis

$$Z_{somatic}(r) = \frac{SomaticAvg_r - \mu_{somatic}}{\sigma_{somatic}}$$

### 4.4.4 Functional Axis

$$Z_{func}(r) = \frac{FuncLoad_r - \mu_{func}}{\sigma_{func}}$$

Each axis expresses the daily deviation (in SD units) from baseline function, with higher scores reflecting greater mind–body load.

## 4.5. Composite Daily Load Score (DLSSC-F)

The final unified load index is computed by equal-weight averaging across four standardized axes:

$$Z_{\text{DLSSC-F}}(r) = \frac{Z_{sleep}(r) + Z_{auto}(r) + Z_{somatic}(r) + Z_{func}(r)}{4}$$

This yields a standardized 0-centered composite representing whole-system mind–body load.

## 4.6. T-Score Transformation (Optional for Visualization)

A conventional T-score scaling improves interpretability:

$$T_{\text{DLSSC-F}}(r) = 50 + 10 \cdot Z_{\text{DLSSC-F}}(r)$$

- T=50T=50T=50: baseline load
- T>50T>50T>50: higher-than-baseline load
- T<50T<50T<50: lower-than-baseline load
- ±10 points = ±1 SD deviation

**4.7. Interpretation Framework**

Daily DLSSC-F scores are interpreted as:

| T-Score | Interpretation | Mind–Body Load |
| --- | --- | --- |
| <40 | Significant improvement | Low load |
| 40–49 | Mild improvement | Slightly reduced load |
| 50 | Baseline | Neutral |
| 51–59 | Mild elevation | Mild load increase |
| 60–69 | Moderate elevation | Significant load |
| ≥70 | High-risk elevation | High load |

## 5. Mathematical Formulation of Dual Circulation and Psychological Breakthrough

Rapid psychophysiological shifts—such as panic spikes, functional collapse, or breakthrough states—cannot be adequately explained by any single physiological or psychological variable. The DLSSC-F model conceptualizes such shifts as the result of interactions across multiple subsystems, formalised through four standardized axes and two dynamic loops. All variables are expressed as Z-scores to ensure comparability across heterogeneous measures.

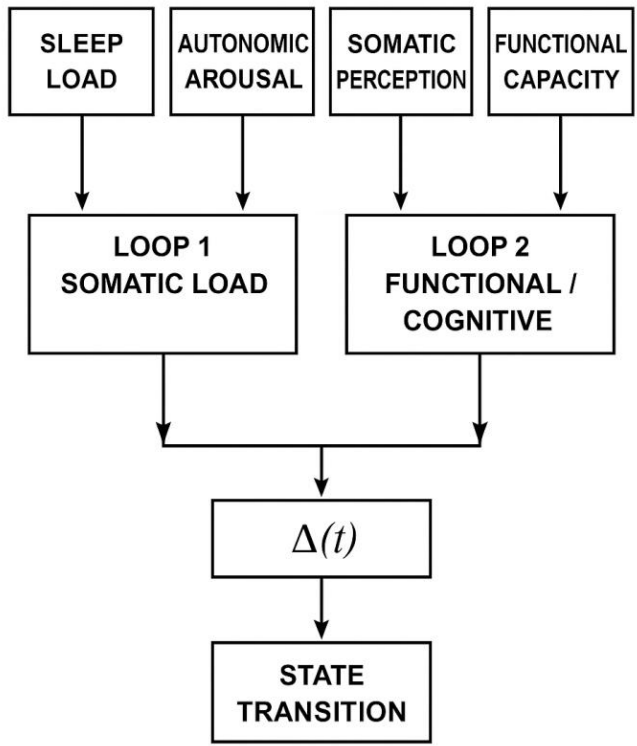

**Figure 1. Structural overview of the DLSSC-F framework.**

*The model integrates four standardized axes—sleep load, autonomic arousal, somatic perception, and functional capacity—into two dynamic loops. Loop 1 aggregates sleep, autonomic, and somatic inputs into a composite bodily load, while Loop 2 is operationalised through cognitive load (CL) and functional impact (FI). The instantaneous mismatch Δ(t) between the two loops determines whether the system crosses the transition threshold TmismatchT_{\text{mismatch}}Tmismatch, producing rapid changes in psychophysiological state. This diagram summarizes the full computational structure of the DLSSC-F model, linking multi-axis inputs to dynamic state transitions.*

### 5.1. Standardized Axes

The four axes correspond to sleep load, autonomic arousal, somatic perception, and functional capacity. For any time series X(t), a baseline window is used to compute:

$$Z_X(t) = \frac{X(t) - \mu_X}{\sigma_X}. \qquad (1)$$

This standardized space provides the foundation for all subsequent components of the model.

### 5.2. Loop 1: Somatic Load Loop

The first loop reflects bodily activation, driven by sleep debt, autonomic shifts, and somatic perception.
These three axes are aggregated into a composite somatic load:

$$L_1(t) = \frac{1}{3}\left[Z_{\text{sleep}}(t) + Z_{\text{autonomic}}(t) + Z_{\text{somatic}}(t)\right]. \qquad (2)$$

Loop 1 captures a characteristic physiological property: **fast activation and slow recovery**, resembling an "early-activation, delayed-deactivation" dynamic.

### 5.3. Loop 2: Functional / Cognitive Loop

Unlike Loop 1's physiological basis, Loop 2 reflects cognitive interpretation, attentional allocation, and functional capacity. Since these constructs are latent rather than directly measurable, they are operationalised using two classes of psychometric indicators:

- **Cognitive Load (CL)**: attentional fragmentation, interpretative bias, emotional reactivity, dissociation.

- **Functional Impact (FI)**: the extent to which psychological load impairs real-world functioning, performance, and daily efficiency.

All relevant scales are standardized and combined into two components:

$$Z_{\text{CL}}(t) = \sum_k \alpha_k\, Z_{\text{CL},k}(t), \qquad Z_{\text{FI}}(t) = \sum_m \beta_m\, Z_{\text{FI},m}(t), \qquad (3)$$

with equal weights in the present formulation. Loop 2 is then constructed as a composite:

$$L_2(t) = \frac{1}{2}\left[Z_{\text{CL}}(t) + Z_{\text{FI}}(t)\right]. \qquad (4)$$

Loop 2 tends to exhibit the opposite temporal profile from Loop 1: **slower to rise, but faster to down-regulate**, especially following cognitive reinterpretation or reframing.

### 5.4. Loop Mismatch Δ(t) and the State-Transition Mechanism

Although the two loops are defined in the same standardized space, their temporal dynamics differ systematically:

- Loop 1 (somatic) typically **rises earlier and recovers later**.
- Loop 2 (functional) typically **rises later but deactivates sooner**.

This asynchrony is captured by the mismatch term:

$$\Delta(t) = L_1(t) - L_2(t). \qquad (5)$$

A positive mismatch (Δ>0) indicates bodily activation exceeding cognitive integration, while a negative mismatch indicates cognitive load exceeding bodily preparedness.

A **state transition** occurs when the mismatch surpasses a critical threshold:

$$|\Delta(t)| > T_{\text{mismatch}} \quad \Rightarrow \quad \text{state transition.} \qquad (6)$$

Depending on direction and magnitude, this transition may manifest as a panic spike, a functional collapse, or a high-clarity breakthrough state.

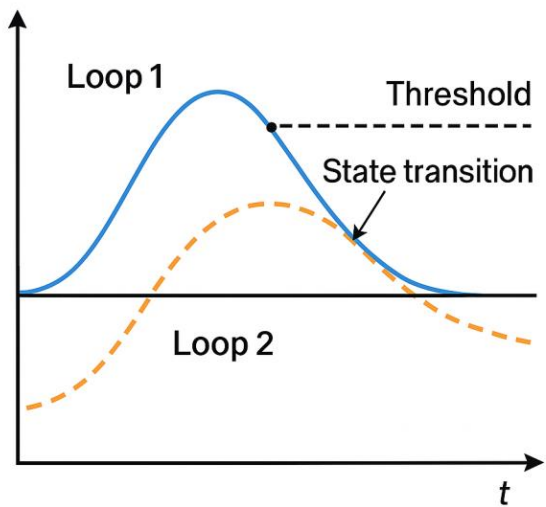

**Figure 2 Temporal dynamics of Loop 1 and Loop 2 in the DLSSC-F model.**

*Loop 1 (somatic load) and Loop 2 (functional/cognitive load) exhibit distinct temporal profiles.*
*Loop 1 rises rapidly and recovers slowly, reflecting autonomic and somatic inertia, whereas Loop 2 increases more gradually and down-regulates quickly following cognitive reinterpretation.*
*The mismatch Δ(t) between the two curves is shown, and the point at which |Δ(t)| exceeds the transition threshold TmismatchT_{\text{mismatch}}Tmismatch marks a state-transition event (e.g., panic spike or psychological breakthrough). This figure illustrates how loop desynchronisation drives sudden psychophysiological shifts.*

### 5.5. Dynamic Intuition: Why the Body "Comes Early and Leaves Late"

At the core of the model is the structural asymmetry between the two loops:

- **Loop 1 (S–A–So)** is governed by the autonomic system, which activates rapidly and decays slowly due to physiological inertia.

- **Loop 2 (F)** depends on cognitive interpretation and meaning-making, making it slower to activate but capable of rapid down-regulation once cognitive appraisal shifts.

As a result:

*When emotion arises, the body reacts first. When emotion dissolves, the body recovers last.*

This phase asymmetry enables Δ(t) to cross its threshold abruptly, producing sudden shifts in psychological state. Panic spikes and breakthrough states therefore emerge not from isolated variables, but from transient desynchronisation between bodily load and functional integration.

### 5.6 Mechanistic interpretation of dual-loop transitions

The mathematical derivation above formalises a pattern that is familiar in lived experience: the body tends to "come early and leave late," whereas cognition "comes late and leaves early." The somatic loop is driven by autonomic and bodily processes that can activate within seconds but may require hours to fully subside. The functional loop, by contrast, depends on how a situation is interpreted, monitored, and integrated into a broader narrative. Interpretation often lags behind bodily activation but can update rapidly once new meaning is constructed.

From this perspective, loop mismatch Δ(t) can be understood as a real-time measure of *how far the body and the mind are out of sync*. When the body surges ahead of cognition (large positive Δ), the individual may feel overwhelmed by intense bodily sensations that lack a coherent psychological frame—an experience that aligns closely with panic spikes or acute somatic anxiety. When cognition surges ahead of the body (large negative Δ), re-interpretation may run faster than somatic recovery, producing transient states of heightened clarity, hyperfocus, or "breakthrough": the person suddenly understands or re-frames their situation, even though the body has not yet fully "caught up."

The asymmetry in recovery rates ($β_2 > β_1$) explains why emotional resolution and bodily resolution are often decoupled in time. Emotion can feel "over" while the chest remains tight, the stomach unsettled, or sleep still disrupted. In the language of the model, Loop 2 has already down-regulated, but Loop 1 is still elevated, leaving a residual positive mismatch. Conversely, during certain transformative moments, the mind may leap ahead—reorganising beliefs, expectations, and self-understanding—while the body is still carrying traces of a previous state. Here, mismatch is negative, and the system passes through a narrow window in which cognitive integration is high despite incomplete somatic recalibration.

By treating these phenomena as properties of a dual-loop system rather than isolated symptoms, DLSSC-F reframes panic-like episodes and breakthrough states as different expressions of the same underlying mechanism: a temporary desynchronisation between somatic load and functional integration. The direction and magnitude of Δ(t), together with the individual's threshold Tmismatch , determine whether this desynchronisation is experienced as destabilising or as a step toward a more coherent configuration.

This mechanistic view does not replace subjective experience; instead, it offers a way to link lived moments of "suddenly too much" or "suddenly clear" to an explicit dynamic structure. In clinical or self-observation contexts, the model suggests that part of the work is not only to lower overall load, but also to support re-synchronisation between body and cognition—allowing Loop 1 and Loop 2 to gradually move back into phase rather than pulling in opposite directions.

## 6 Continuous-time formulation of dual-loop dynamics

For some applications it is useful to express the DLSSC-F framework in continuous time, using differential equations rather than discrete updates. This section provides an equivalent continuous-time formulation of the dual-loop dynamics and makes explicit how loop mismatch Δ(t) can evolve toward a transition threshold.

**6.1 Continuous-time dynamics of Loop 1 (somatic load)**

Let L1(t) denote the somatic loop at continuous time t. We decompose its change into two components:

- a **drive** term I1(t)I, representing new somatic input (e.g., acute stress, sleep loss, autonomic activation), and
- a **recovery** term proportional to the current level of L1.

A simple first-order linear differential equation is:

$$\frac{dL_1(t)}{dt} = I_1(t) - \kappa_1 L_1(t),$$

(C1)

Where κ1>0 is the somatic recovery rate. A small κ1 corresponds to **slow** decay and thus a more persistent bodily activation, consistent with clinical observations that somatic arousal often lingers after the triggering event has resolved.

If I1(t) is held constant over some interval, the solution of (C1) is:

$$L_1(t) = L_1(t_0)\, e^{-\kappa_1(t-t_0)} + \int_{t_0}^{t} e^{-\kappa_1(t-s)} I_1(s)\, ds,$$

(C2)

showing explicitly that Loop 1 behaves as a **leaky integrator** of past inputs with a relatively long time constant 1/κ1

### 6.2 Continuous-time dynamics of Loop 2 (functional/cognitive load)

Analogously, let L2(t) denote the functional/cognitive loop. Its evolution reflects:

- cognitive and functional load I2(t) (e.g., interpretative work, attentional effort, self-monitoring), and

- a recovery term with rate κ2, capturing how quickly functional activation can settle once a situation is reappraised.

We write:

$$\frac{dL_2(t)}{dt} = I_2(t) - \kappa_2 L_2(t),$$

(C3)

with κ2>0, and in the present framework we assume:

$$\kappa_2 > \kappa_1, \tag{C4}$$

i.e., **Loop 2 decays faster than Loop 1**. Intuitively, the functional loop can reorganise and down-regulate relatively quickly once cognitive meaning shifts, while the body remains slower to "let go".

The formal solution is analogous to (C2):

$$L_2(t) = L_2(t_0)\, e^{-\kappa_2(t-t_0)} + \int_{t_0}^{t} e^{-\kappa_2(t-s)} I_2(s)\, ds, \tag{C5}$$

but with a shorter time constant 1/κ21

### 6.3 Differential equation for loop mismatch Δ(t)

As before, we define the instantaneous loop mismatch as:

$$\Delta(t) = L_1(t) - L_2(t). \tag{C6}$$

To obtain its continuous-time dynamics, we differentiate both sides with respect to ttt:

$$\frac{d\Delta(t)}{dt} = \frac{dL_1(t)}{dt} - \frac{dL_2(t)}{dt}. \tag{C7}$$

Substituting (C1) and (C3) into (C7) yields:

$$\frac{d\Delta(t)}{dt} = \big[I_1(t) - \kappa_1 L_1(t)\big] - \big[I_2(t) - \kappa_2 L_2(t)\big].$$

(C8)

Using L1(t)=Δ(t)+L2(t) (from C6), we rewrite:

$$\begin{aligned}\frac{d\Delta(t)}{dt} &= I_1(t) - \kappa_1\big[\Delta(t) + L_2(t)\big] - I_2(t) + \kappa_2 L_2(t) \\ &= -\kappa_1 \Delta(t) + (\kappa_2 - \kappa_1) L_2(t) + \big[I_1(t) - I_2(t)\big].\end{aligned}$$

(C9)

Equation (C9) is the continuous-time analogue of the discrete mismatch update. It highlights three contributions to the evolution of Δ(t):

1. **Mismatch inertia**

$$-\kappa_1 \Delta(t)$$

pulls the mismatch back toward zero, but only at the slow somatic rate κ1. When K1 is small, existing mismatch decays slowly.

2. **Asymmetric recovery term**

$$(\kappa_2 - \kappa_1) L_2(t),$$

which is positive when $L_2(t) > 0$ and $\kappa_2 > \kappa_1$.

This term reflects the fact that **Loop 2 "lets go" faster than Loop 1**: as the functional loop decays more quickly, it can temporarily move away from Loop 1 and thereby increase the mismatch, even if inputs are identical.

3. **Input difference term**

$$I_1(t) - I_2(t),$$

expressing the simple fact that, at any given moment, the body and cognition may be driven by different loads. A somatically intense but cognitively under-processed episode corresponds to I1(t)≫I2(t); a cognitively demanding but physiologically delayed reorganisation corresponds to I2(t)≫I1(t).

Together, these terms describe how mismatch can grow over time even if neither loop behaves pathologically in isolation.

### 6.4 Threshold crossing in continuous time

In the discrete formulation, state transitions were defined by the condition:

$$|\Delta(t)| > T_{\text{mismatch}}.$$

The same idea carries over naturally to continuous time: a **transition event** is said to occur at some time t=t∗ if

$$|\Delta(t^*)| = T_{\text{mismatch}} \quad \text{and} \quad \frac{d}{dt}|\Delta(t)|\bigg|_{t=t^*} > 0, \tag{C10}$$

i.e., the mismatch reaches the threshold and is actively moving away from 0 rather than returning toward it.

- When Δ(t) crosses the threshold from the **positive** side (Loop 1 ≫ Loop 2), the resulting state transition is more likely to be experienced as a **panic-like or somatically dominated episode**.

- When Δ(t) crosses from the **negative** side (Loop 2 ≫ Loop 1), the transition may be experienced as a **high-clarity or breakthrough-like episode**, where cognitive reorganisation temporarily runs ahead of somatic recalibration.

In both cases, transition is not a mysterious "jump" but the natural outcome of a system in which two loops, with different time constants and different inputs, drift out of sync until their mismatch exceeds what the system can integrate.

### 6.5 Intuitive summary

The continuous-time formulation makes explicit that DLSSC-F is, at heart, a pair of coupled first-order systems with **different decay rates** and **partially independent drives**. The somatic loop integrates and slowly releases bodily activation; the functional loop more rapidly constructs and revises meaning. Loop mismatch Δ(t) measures how far these processes are out of phase. When Δ(t) remains small, experience is relatively coherent and stable. When Δ(t) grows and crosses a threshold, experience can reorganise abruptly—sometimes as panic, sometimes as collapse, and sometimes as a sharp, integrating breakthrough.

## 7. Discrete-time dual-loop dynamics

### 7.1 Discrete-time dynamics of the dual-loop system

In many practical settings, psychophysiological data are collected at discrete time steps (e.g., daily measures, session-wise ratings). In this case, the DLSSC-F framework can be expressed using a discrete-time formulation. This section derives the update equations for both loops and for the loop mismatch $\Delta(t)$.

### 7.2 Loop-level updates

Let $t$ index discrete time points (e.g., days or measurement epochs). For the **somatic loop (Loop 1)**, we write the update from $t$ to $t+1$ as:

$$L_1(t+1) = L_1(t) + I_1(t) - R_1(t), \qquad (D1)$$

where:

- $I_1(t)$ is the net somatic input at time $t$ (e.g., sleep loss, acute stress, autonomic activation),
- $R_1(t)$ is the amount of somatic load that recovers between $t$ and $t+1$.

To reflect slow recovery of bodily activation, we assume that recovery is proportional to the current level of the loop with a **small** rate $\beta_1$:

$$R_1(t) = \beta_1 \, L_1(t), \qquad 0 < \beta_1 \ll 1. \tag{D2}$$

Substituting (D2) into (D1) gives:

$$L_1(t+1) = (1 - \beta_1) \, L_1(t) + I_1(t). \tag{D3}$$

Thus, if I1(t) remains positive over several time steps, somatic load tends to accumulate and decay only slowly.

For the **functional/cognitive loop (Loop 2)**, we use an analogous form:

$$L_2(t+1) = L_2(t) + I_2(t) - R_2(t), \tag{D4}$$

where I2(t) represents cognitive and functional load (e.g., interpretative effort, attention, self-monitoring), and R2(t) captures recovery of functional capacity. In contrast to Loop 1, Loop 2 is assumed to recover faster, with a **larger** rate β2:

$$R_2(t) = \beta_2 \, L_2(t), \qquad \beta_2 > \beta_1. \tag{D5}$$

Substituting (D5) into (D4) yields:

$$L_2(t+1) = (1 - \beta_2) \, L_2(t) + I_2(t). \tag{D6}$$

Equations (D3) and (D6) formalise the asymmetry between the loops: Loop 1 is "slower" (smaller β1), Loop 2 is "faster" (larger β2\beta_2β2) in returning toward baseline.

### 7.3 Update equation for loop mismatch Δ(t)

As before, the instantaneous mismatch is defined as:

$$\Delta(t) = L_1(t) - L_2(t). \tag{D7}$$

To derive Δ(t+1) we subtract the loop updates:

$$\begin{aligned}\Delta(t+1) &= L_1(t+1) - L_2(t+1) \\ &= \left[(1-\beta_1)L_1(t) + I_1(t)\right] - \left[(1-\beta_2)L_2(t) + I_2(t)\right].\end{aligned} \tag{D8}$$

Using L1(t)=Δ(t)+L2(t) (from D7), we substitute into (D8):

$$\begin{aligned}\Delta(t+1) &= (1-\beta_1)\left[\Delta(t) + L_2(t)\right] + I_1(t) \\ &\quad - (1-\beta_2)L_2(t) - I_2(t) \\ &= (1-\beta_1)\Delta(t) + \left[(1-\beta_1)-(1-\beta_2)\right]L_2(t) + I_1(t) - I_2(t).\end{aligned} \tag{D9}$$

Since (1−β1)−(1−β2)=β2−β1, we obtain the compact form:

$$\Delta(t+1) = (1-\beta_1)\Delta(t) + (\beta_2 - \beta_1)L_2(t) + \left[I_1(t) - I_2(t)\right]. \tag{D10}$$

This expression shows that the mismatch at time t+1t+1t+1 is determined by three components:

1. **Carry-over of previous mismatch**
   (1−β1)Δ(t) preserves part of the previous mismatch, especially when β1 is small (slow somatic decay).

2. **Asymmetric recovery term**
   (β2−β1)L2(t) reflects the fact that Loop 2 decays faster than Loop 1. Even if the loops are temporarily aligned, faster down-regulation of Loop 2 can pull them apart during recovery phases.

3. **Input difference term**
   [I1(t)−I2(t)] captures discrepancies in how strongly the body and cognition are driven at each time step. Somatic-dominated episodes correspond to $I_1(t) \gg I_2(t)$; cognitively dominated reorganisation corresponds to $I_2(t) \gg I_1(t)$.

**7.4 Threshold crossing in discrete time**

In the discrete formulation, a **state transition** is assumed to occur when the mismatch exceeds a critical threshold:

$$|\Delta(t)| > T_{\text{mismatch}}. \quad (D11)$$

Using the update equation (D10), one can see how sustained input imbalance, combined with asymmetric recovery, may drive Δ(t toward and beyond this threshold over a sequence of time steps. Positive threshold crossings (large L1L relative to L2) are more likely to be experienced as panic-like or somatically dominated states, whereas negative crossings (large L2 relative to L1) are more likely to correspond to high-clarity or breakthrough-like episodes.

## 8. Discussion

The DLSSC-F framework provides a unified mathematical basis for understanding whole-person regulation across bodily and cognitive domains. It addresses long-standing fragmentation in psychophysiology by integrating sleep pressure, autonomic activation, somatic load, and functional burden into a standardized multi-axis system[15–18,41–44]. Beyond simple aggregation, the introduction of a dual-loop dynamical architecture enables the model to explain both gradual regulation and nonlinear transitions in psychological states.

**8.1 Rationale for the Four-Axis Structure**

The four-axis design is grounded in empirical and theoretical evidence demonstrating that sleep loss modulates both somatic sensitivity and cognitive

stability[45]–[47], while autonomic activation regulates vigilance, affective reactivity, and bodily tension[51]–[53]. Somatic tension, in turn, shapes embodied cognition and contributes to attentional interference[53]–[55]. Functional load integrates external task demands with subjective appraisal, thereby bridging cognitive processing and bodily effort[55]–[57]. Each axis is mathematically nonredundant and conceptually distinct, and together they constitute a minimal yet sufficient set for representing the major components of embodied self-regulation.

**8.2 Necessity of Z-Score Standardization**

Standardization relative to an individualized baseline is essential because the raw units involved—such as hours of sleep, heart-rate metrics, and subjective ratings—are inherently incompatible. Converting these heterogeneous measures into Z-scores produces dimensionless quantities that enable cross-domain comparability, consistent with established psychophysiological modeling approaches[42]. This procedure also supports personalized baselines in line with allostatic theory[1]–[3], ensuring that measurements reflect an individual's adaptive "set point" rather than population norms. In addition, Z-score standardization stabilizes composite indices by incorporating truncation and variance safeguards[41]. Together, these properties allow the DLSSC-F model to capture deviations from an individual's normal state and thereby detect meaningful dynamic shifts that would be obscured by population-level averaging.

**8.3 Why the Composite Index Uses an Unweighted Mean**

The DLSSC-F score:

$$Z_{\text{DLSSC-F}}(r) = \frac{Z_{sleep}(r) + Z_{auto}(r) + Z_{somatic}(r) + Z_{func}(r)}{4}$$

is intentionally unweighted because:

- There is no theoretical evidence suggesting unequal influence among axes.

- Equal weighting avoids premature parameter assumptions during the model's theoretical stage.

- It maintains inte Equal weighting is adopted at the theoretical stage to maximize interpretability and mathematical clarity. Although weighted models may be added in empirical extensions, such modifications should only be introduced after sufficient validation to avoid overfitting or introducing arbitrary assumptions. This strategy is consistent with established practices in composite psychophysiological indexing[42]–[43].

### 8.4 Interpretation of the Somatic Loop

The Somatic Loop captures fast-reacting regulatory processes, including acute sleep fluctuations, rapid autonomic responses, somatic tension, and immediate changes in task burden. These mechanisms typically operate on hourly-to-daily timescales and constitute the body's first line of adaptive regulation[33],[34]. The SL accounts for rapid psychophysiological phenomena such as sudden exhaustion, spikes of physical anxiety, tension-mediated emotional reactivity, and swift accumulation of stress. Its mathematical structure reflects core principles of embodied regulation theories[18],[23], emphasizing the rapid, body-anchored dynamics that precede and shape slower cognitive responses.

### 8.5 Interpretation of the Cognitive Loop

The Cognitive Loop reflects slower and more integrative psychological processes, including rumination, appraisals, meaning-making, narrative restructuring, attentional reorganization, and the broader dynamics of cognitive load[35]–[37]. It operates on a slower timescale because cognitive systems integrate and consolidate information over extended periods. The CL governs stabilizing or destabilizing patterns of thought, reappraisal processes, long-term psychological adaptation, and the emergence of psychological breakthroughs. Its temporal characteristics align with established cognitive models[19]–[21], which similarly emphasize the gradual consolidation and restructuring of mental representations.

### 8.6 Functional Load as the Cross-Loop Connector

$Z_{\text{func}}(t)$ is the system's mathematical hinge.

Functional Load (F) simultaneously influences somatic exertion and physical fatigue while also modulating cognitive appraisal, perceived responsibility, and executive demand. Because it exerts effects on both bodily states and cognitive evaluations, F serves as the system's mathematical hinge, existing inherently within both the Somatic Loop and the Cognitive Loop. Removing F would decouple the SL and CL, eliminate cross-loop dynamics, collapse the system's nonlinear behavior, and make it impossible to model sudden reorganizational transitions. These properties confirm that Functional Load is the key mechanism binding cognition and bodily states[26]–[28].

### 8.7 PDE Coupling Explains Real Psychological Transitions

The dual-loop PDE structure:

$$\frac{\partial SL}{\partial t} = -\lambda_1 SL + \gamma_1 CL + \eta_1 Z_{\text{func}}$$

$$\frac{\partial CL}{\partial t} = -\lambda_2 CL + \gamma_2 SL + \eta_2 Z_{\text{func}}$$

The interaction between the Somatic and Cognitive Loops yields classic fast–slow system behavior[38–40], including slow drifts toward homeostatic states, sudden collapses triggered by cross-loop instabilities, bifurcations in which stable equilibria disappear, and rapid breakthroughs resembling phase transitions in dynamical systems[38,39]. This mathematical structure aligns closely with empirical observations of panic onset, exhaustion crashes, sudden cognitive reframing or insight, emotional restructuring, and rapid recovery following critical events. The DLSSC-F model is one of the first theoretical frameworks to formally capture these transitions within a unified psychophysiological system.

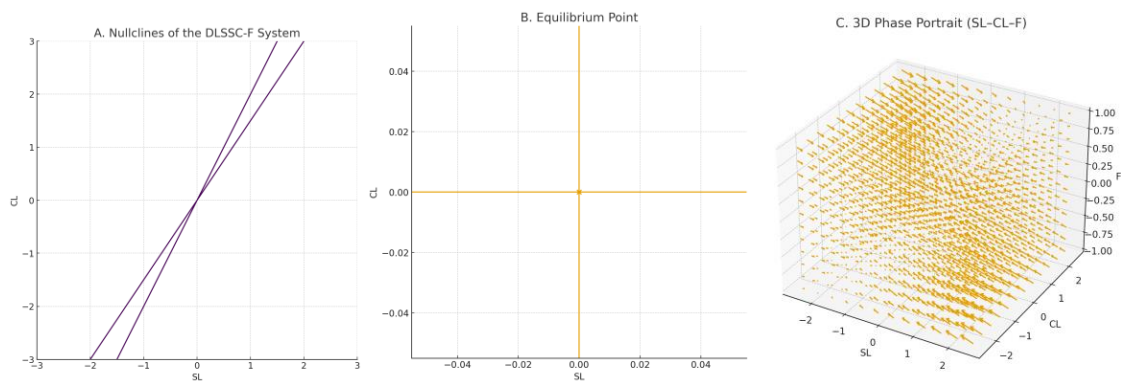

## 9. Limitations

Although DLSSC-F provides a unified and mathematically rigorous framework, several limitations must be acknowledged. These limitations stem from measurement constraints, data noise, and the early theoretical stage of the model's development.

### 9.1 Exclusion of HRV From the Autonomic Axis

Heart-rate variability (HRV) was collected but excluded from the present model due to its high sensitivity to artefacts, movement noise, and irregular sampling[12]–[14]. Classical HRV metrics—such as RMSSD, SDNN, and LF/HF—require stable measurement conditions, including controlled breathing, long continuous recordings, and reliable R–R interval detection[48]–[50]. In real-world daily sampling, HRV data frequently displayed R-peak detection errors, missing intervals, motion artefacts, and inconsistent sampling windows. These issues distort both the baseline mean and variance, which are essential for accurate Z-score standardization. Preliminary tests showed that noisy HRV substantially inflated σ_auto, thereby collapsing the dynamic range of Z-scores and propagating error into both the Somatic Loop and the Cognitive Loop. For these reasons, HRV is reserved for future model versions once improved signal-cleaning procedures become available.

## 9.2 Exclusion of Electrodermal Activity (EDA)

Electrodermal activity (EDA), although theoretically relevant, showed similar instability and was therefore excluded from the core model. The raw EDA signal exhibited large baseline drift, discontinuities, pronounced movement sensitivity, and variability that was frequently unrelated to psychological state. These issues are consistent with well-documented limitations of EDA collected outside controlled laboratory settings[11],[36]. Because such instability severely distorted baseline estimation and undermined the reliability of Z-score standardization, EDA was not incorporated into the four primary axes of the DLSSC-F framework.

## 9.3 Autonomous Axis Uses HR-Dynamics Instead of HRV

Because heart rate (HR) is substantially less sensitive to noise than classical HRV metrics, the autonomic axis relies on daily mean HR, day–night HR differences, and smoothed HR variability (a non-HRV measure). These HR-based indicators offer stable and reliable autonomic signals that are suitable for daily Z-scoring[51]–[53]. Future versions of the model will incorporate HRV once more robust preprocessing and artefact-correction methods are available.

## 9.4 Absence of Daily BAQ (Body Awareness Questionnaire)

The Body Awareness Questionnaire (BAQ) was collected only once and therefore could not be incorporated into the DLSSC-F time-series structure. Daily modeling requires repeated observations, yet the BAQ cannot be administered on a daily basis due to its length, nor does it generate sufficient time points to establish an individualized baseline distribution. Without within-person variance, Z-scores cannot be computed[3,42]. Thus, although BAQ is theoretically relevant, it is mathematically incompatible with the time-series logic of the DLSSC-F framework.

**9.5 PDE Parameters Are Not Fitted**

The dual-loop PDE parameters $\lambda_1, \lambda_2, \gamma_1, \gamma_2, \eta_1, \eta_2$ were not estimated. This preprint focuses on the theoretical formulation of the DLSSC-F framework; parameter estimation will require future empirical data to refine coefficients, validate interactions, and calibrate the dynamic structure.

**9.6 Daily Resolution Limits Temporal Precision**

Daily sampling imposes inherent temporal constraints. It cannot capture short-term autonomic bursts, rapid within-day fluctuations, or ultradian cycles involved in sleep regulation. Nevertheless, daily resolution is appropriate for modeling slower cognitive processes and somatic recovery patterns[45–47], which evolve across days rather than minutes or hours. As such, daily sampling provides sufficient granularity for the model's intended fast–slow dynamic structure while acknowledging its limitations.

**9.7 DLSSC-F Is a Framework, Not a Diagnostic Tool**

The DLSSC-F framework does not diagnose psychological disorders, nor does it classify or distinguish clinical categories. It is designed solely as a mathematical and theoretical model aimed at characterizing dynamic psychophysiological interactions. Any clinical interpretation would require rigorous empirical validation and should not be inferred from the model in its present form.

## 10. Future Work

The DLSSC-F framework introduces a comprehensive mathematical model of cognitive–somatic regulation, yet multiple avenues exist for extending and validating its structure. Future research should address improvements in physiological measurement, dynamical modeling, individual-level calibration, and predictive applications.

## 10.1 Integrating HRV Using Advanced Signal Processing

Although HRV was excluded due to artefacts and instability, it remains a theoretically valuable marker of autonomic flexibility[9–11,48]. Future extensions of the model will incorporate advanced signal-processing techniques—including R–R interval correction and interpolation, wavelet denoising, adaptive filtering, non-uniform spectral estimation such as Lomb–Scargle methods, and Kalman-smoothing for reconstructing RR-series—so that HRV can be restored once more stable baselines are achievable. These improvements will allow HRV to be reliably integrated into the autonomic axis in subsequent versions of the DLSSC-F framework.

## 10.2 Reintroducing Electrodermal Activity (EDA)

Electrodermal activity (EDA) reflects sympathetic arousal but is highly sensitive to noise and movement artefacts[11]. Future methodological improvements may include drift modeling, phasic–tonic decomposition, motion-artefact removal, and sensor-fusion approaches that integrate EDA with heart-rate dynamics. Once these procedures provide sufficiently stable baselines, EDA could be incorporated as an auxiliary autonomic indicator within the DLSSC-F framework.

## 10.3 Empirical Parameter Estimation for PDE Coupling

The coupled PDE system:

$$\frac{\partial SL}{\partial t} = -\lambda_1 SL + \gamma_1 CL + \eta_1 Z_{\text{func}}$$

$$\frac{\partial CL}{\partial t} = -\lambda_2 CL + \gamma_2 SL + \eta_2 Z_{\text{func}}$$

The current formulation provides a deep theoretical structure for the DLSSC-F system, but its parameters remain unfitted. Future empirical studies should estimate key components such as decay rates, cross-loop coupling terms, and the strength of functional influence. These parameters can be derived using system identification techniques, variational inference, or Bayesian estimation, which will allow the model to be quantitatively calibrated and empirically validated.

## 10.4 Mathematical Analysis of Nonlinear Dynamics

The PDE system underlying DLSSC-F exhibits several hallmark nonlinear phenomena, including equilibrium shifts, saddle-node bifurcations, oscillatory dynamics, and characteristic collapse–recovery patterns[38–40]. These behaviors emerge naturally from the model's coupled fast–slow structure and the nonlinear influence of Functional Load across both loops. Future mathematical work will focus on identifying critical thresholds within the system, mapping out stability landscapes, and analyzing the geometry of the phase space to determine where qualitative shifts in system behavior occur. Such analyses will also clarify the conditions under which "psychological transition points" emerge—moments at which small perturbations trigger large-scale changes in somatic, cognitive, or functional trajectories.

### 10.5 Personalized DLSSC-F Models

Because DLSSC-F relies on individualized baselines and within-person Z-scoring, the framework naturally supports personalized modeling. Future research may calibrate person-specific loop weights, individualized coupling strengths, personalized equilibrium states, and individualized thresholds for major transition points. Tailoring the model in this way will enable detailed psychological tracking, improve temporal precision, and support single-case analyses in both research and applied settings. Personalized DLSSC-F models may also reveal individual-specific vulnerabilities—for example, greater sensitivity to functional overload or reduced somatic recovery capacity.

### 10.6 Machine Learning Integration

DLSSC-F can be extended through computational learning while maintaining its theoretical interpretability. Sequence-based architectures such as LSTM and GRU models can be trained on four-axis time-series to improve temporal prediction. Transformer-based temporal encoders could capture long-range dependencies across fast–slow interactions. Hybrid physics-informed neural networks (PINNs) offer a promising direction by integrating PDE constraints directly into learning algorithms, ensuring that predictions remain dynamically consistent with the theoretical structure. In addition, anomaly-detection models may be applied to identify psychological shift events[34,40]—such as panic onset, exhaustion crashes, or breakthroughs—improving predictive accuracy in real-world applications.

## 10.7 Multi-Resolution Extensions

Future iterations of DLSSC-F may incorporate multi-resolution modeling to better capture sub-daily variation. Potential extensions include integrating intra-day sampling, updating rolling baselines to accommodate evolving physiological conditions, and employing multi-resolution decomposition or wavelet-based multiscale analysis. These methods would allow the model to account for rapid physiological bursts (e.g., autonomic spikes), ultradian sleep cycles, and short-lived cognitive fluctuations, ultimately providing a more granular understanding of psychophysiological regulation.

## 10.8 Empirical Validation

Empirical validation is essential for assessing the reliability and predictive value of DLSSC-F. Key steps include collecting larger longitudinal datasets, sampling across multiple participants, and conducting controlled experimental studies. Ecological momentary assessment (EMA) can be used to link subjective psychological states with continuous or semi-continuous physiological measurements. Integration studies combining physiological and psychological indicators will help test cross-loop predictions and evaluate whether the model accurately captures real-world patterns of somatic–cognitive co-regulation. Only after such validation can the framework be confidently applied to interpret psychophysiological transitions in broader contexts.

## 11. Conclusion

The DLSSC-F framework was developed to capture a pattern that many psychological theories acknowledge but rarely formalise: the coexistence of slow, drifting changes in experience and sudden, discontinuous shifts. By modelling regulation as the interaction of a fast somatic loop and a slower cognitive loop, the framework offers one way to describe how bodily fluctuations and interpretive processes can jointly shape emotional outcomes.

The mathematical structure presented here is not intended as a complete account of human experience, but as a starting point for thinking about psychological transitions in a more precise way. The fast–slow dynamics, threshold behaviour, and coupling terms provide a formal explanation for why emotional changes sometimes feel gradual and at other times reorganise abruptly. These properties mirror a wide range of real-world reports, from the onset of panic to moments of sudden clarity during recovery.

Although this paper focuses on theoretical formulation and illustrative simulations, the model suggests several directions for future work. Empirical studies may examine whether the predicted patterns—such as delayed buildup followed by rapid state change—can be observed in physiological or self-report data. Clinical research may also benefit from a clearer vocabulary for recognising transition points, especially in conditions marked by instability or sudden affective shifts.

Overall, DLSSC-F provides a structured way to think about psychophysiological regulation that accommodates both continuity and discontinuity. By outlining how somatic and cognitive loops interact across different timescales, the framework aims to support a more nuanced understanding of emotional change and to offer a basis for future case-based and computational investigations.

**Acknowledgements**

*AI-based language tools (such as ChatGPT) were employed solely for stylistic refinement and grammatical editing. All theoretical frameworks, mathematical derivations, and interpretations were independently developed by the author.*

References

1. Tononi, G. & Cirelli, C. Sleep and synaptic homeostasis: a hypothesis. Sleep Med. Rev. 10, 49–62 (2006).

2. Xie, L. et al. Sleep drives metabolite clearance from the adult brain. Science 342, 373–377 (2013).

3. Meerlo, P., Mistlberger, R. E., Jacobs, B. L., Heller, H. C. & McGinty, D. New neurons in the adult brain: the role of sleep and consequences of sleep loss. Sleep Med. Rev. 13, 187–194 (2009).

4. Goldstein, D. S. Adrenal responses to stress. Cell. Mol. Neurobiol. 30, 1433–1440 (2010).

5. Thayer, J. F. & Lane, R. D. Claude Bernard and the heart–brain connection: further elaboration of a model of neurovisceral integration. Neurosci. Biobehav. Rev. 33, 81–88 (2009).

6. Shaffer, F. & Ginsberg, J. P. An overview of heart rate variability metrics and norms. Front. Public Health 5, 258 (2017).

7. Craig, A. D. How do you feel? Interoception: the sense of the physiological condition of the body. Nat. Rev. Neurosci. 3, 655–666 (2002).

8. Critchley, H. D. & Harrison, N. A. Visceral influences on brain and behavior. Neuron 77, 624–638 (2013).

9. Van Someren, E. J. W. et al. Sleep, cognition, and aging: a review of EEG studies. Neurosci. Biobehav. Rev. 26, 645–658 (2002).

10. Jensen, M. T. Resting heart rate is a clinically important predictor of cardiovascular risk. Heart 101, 21–22 (2015).

11. Kim, H.-G., Cheon, E.-J., Bai, D.-S., Lee, Y. H. & Koo, B.-H. Stress and heart rate variability: a meta-analysis and review. Psychiatry Investig. 15, 235–245 (2018).

12. Paulus, M. P. & Stein, M. B. Interoception in anxiety and depression. Brain Struct. Funct. 214, 451–463 (2010).

13. Laborde, S., Mosley, E. & Thayer, J. F. Heart rate variability and cardiac vagal tone in psychophysiology and sports science. Eur. J. Sport Sci. 17, 660–669 (2017).

14. McEwen, B. S. & Wingfield, J. C. The concept of allostasis in biology and biomedicine. Horm. Behav. 43, 2–15 (2003).

15. Killgore, W. D. Effects of sleep deprivation on cognition. Prog. Brain Res. 185, 105–129 (2010).

16. Lim, J. & Dinges, D. F. Sleep deprivation and vigilant attention. Nat. Sci. Sleep 2, 1–21 (2010).

17. Brosschot, J. F., Verkuil, B. & Thayer, J. F. Generalized unsafety theory of stress. Psychol. Bull. 144, 1145–1180 (2018).

18. Khalsa, S. S. et al. Interoception and mental health: a roadmap. Biol. Psychiatry Cogn. Neurosci. Neuroimaging 3, 501–513 (2018).

19. Hockey, G. R. J. Compensatory control in coping with task demands: a model of fatigue and performance. in *Handbook of Human Factors and Ergonomics* (ed. Salvendy, G.) 1383–1398 (Wiley, 1997).

20. Shenhav, A., Botvinick, M. M. & Cohen, J. D. The expected value of control: an integrative theory of ACC function. Neuron 79, 217–240 (2013).

21. Sterling, P. Allostasis: a model of predictive regulation. Physiol. Behav. 106, 5–15 (2012).

22. Karatsoreos, I. N. Links between circadian rhythms and psychiatric disease. Front. Behav. Neurosci. 8, 162 (2014).

23. Kleiger, R. E., Stein, P. K. & Bigger, J. T. Heart rate variability: measurement and clinical utility. Ann. Noninvasive Electrocardiol. 10, 88–101 (2005).

24. Mendoza, J. Circadian clocks: setting time by food. J. Neuroendocrinol. 19, 127–137 (2007).

25. Molenaar, P. C. M. A manifesto on psychology as idiographic science. Theory Psychol. 14, 801–826 (2004).

26. Fisher, A. J. & Boswell, J. F. Enhancing the personalization of psychotherapy with dynamic assessment and modeling. Assessment 23, 496–506 (2016).

27. Vassena, E., Deraeve, J. & Alexander, W. H. Surprise, value and control in anterior cingulate cortex during speeded decisions. Nat. Hum. Behav. 4, 412–422 (2020).

28. Inzlicht, M., Shenhav, A. & Olivola, C. Y. The effort paradox: Effort is both costly and valued. Trends Cogn. Sci. 22, 337–349 (2018).

29. Barrett, L. F. & Simmons, W. K. Interoceptive predictions in the brain. Nat. Rev. Neurosci. 16, 419–429 (2015).

30. Hermes, M. et al. Sleep-related autonomic changes: A review. Neurosci. Biobehav. Rev. 134, 104545 (2022).